\documentclass[review]{elsarticle}
\usepackage[utf8x]{inputenc}
\usepackage{lineno,hyperref}
\modulolinenumbers[5]
\usepackage{amsmath}
\usepackage{mathtools}
\journal{Future Generation Computer Systems}

\begin{document}

\begin{frontmatter}

\title{Voronoi diagrams for the distributed sensor network system data processing \tnoteref{mytitlenote}}
\tnotetext[mytitlenote]{Link to the journal \href{https://www.elsevier.com/journals/future-generation-computer-systems/0167-739X/guide-for-authors}{Future Generation Computer Systems}.}

\author{Kondybayeva Almagul}
\author{Giovanna Di Marzo}
\address{Carouge 1212, Geneve}


\author[mysecondaryaddress]{Centre Universitaire d'Informatique, Universite de Geneve\corref{mycorrespondingauthor}}
\cortext[mycorrespondingauthor]{Corresponding author: Kondybayeva Almagul}
\ead{alma.kond@gmail.com}

\address[mymainaddress]{Carouge, Universite de Geneve, Switzerland}

\begin{abstract}
This article represents the computational model for spacial addresation of the sensors in the dynamically changing real-time internet of things system. The model bases on the Voronoi diagrams as a basic data structure.
Problem - the correct data addresation without time delays in real-time processing and database indexation in distributed storages. 
Goal - to develop the real-time processing model of the object location identification in the Voronoi diagram data structure.
Relevance - the research and development on the contemporary issues on the convergence (the N limit up to which the model presents the algorithm convergence), time-delay, correct indexation in the database transactions and adressation throughout wireless radio frequencies in the distribiuted sensor systems.
Solution proposes the Voronoi diagram computational geometry data structure and the sweeping curve algorithm.
Methods represents the following steps: simulation of the dynamically changing agent system using set of points that are based on a contemporary paths of the public transport routes, bykes, vehicles circulations; 3D map rendering on the SITG Canton of Geneva map; Voronoi diagrams calculation with the Fortune's algorithm. 
Results - this work presents 2D static and dynamic Fortune's realization of the Voronoi diagrams, the architecture of the model for constructing a distributed sensor system based on Voronoi diagrams is described.
Scope - geographic charts showing the distribution of parameters, determination of the nearest points from the object, spacial arrangement of objects in nature, solving the problem of finding all nearest neighbors to the object,
database indexing and search transactions based on quandrant trees.
Conclusions - the research shows the great potential of the new data structures based on a class of computational geometry data structures and sweeping curves in spacial distributed dimensions.
\end{abstract}

\begin{keyword}
\texttt{sensor networks}\sep \texttt{computational geometry}\sep \texttt{internet of things} \sep \texttt{Voronoi diagrams} \sep   \texttt{spacial join} \sep \texttt{spacial index}\sep 
\texttt{sweeping curves}\sep \texttt{Morton code}\sep \texttt{Lebesgue curve} \sep \texttt{search operations} \sep   \texttt{distributed database} \sep \texttt{publisher subscriber}
\MSC[2010] 68-06\sep  68P05
\end{keyword}

\end{frontmatter}

\linenumbers

\section{Introduction}
\begin{figure}[th]
	\begin{center}
		\includegraphics[width=\textwidth]{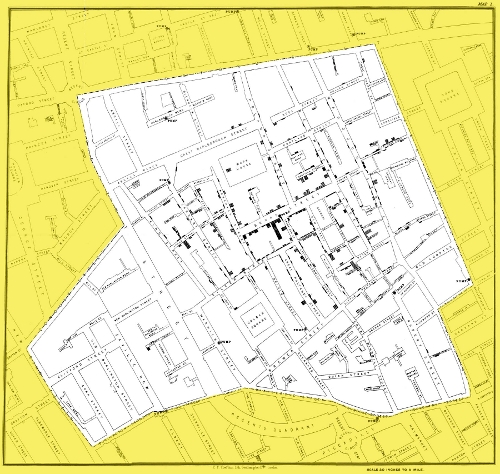}
	\end{center}
	\caption{The area most severely affected by cholera}
	\label{fig:example0}
\end{figure}

Historically, noteworthy applications of the Voronoi diagram or Tissen's polygons emerged in the writings of Jon Snow, widely regarded as the father of modern epidemiology, who studied the cholera that plagued London in 1854
At that time, neither the etiology (note. Cause of occurrence), nor the mode of transmission of this disease were precisely known, there were disputes about two possibilities: infection through contact with the patient, his clothes and / or property, and the theory that the disease spread through atmospheric phenomena such as wind. Snow, using a geographical method, revealed that the cause of the disease was the consumption of water contaminated with feces. To do this, he mapped the distribution of cholera deaths. Then he studied the location of drinking water sources in the city and identified the Voronoi regions for each of them. After calculating the distance between the residence of each victim and the nearest pumping station, he concluded that the area most severely affected by cholera corresponds to the Voronoi area associated with the pumping station on Broad street, as 73 of 83 people died here.
After removing the handle of this pumping station, the outbreak of cholera was extinguished.
The Figure~\ref{fig:example0} represents that map.

\paragraph{Systematization of the main problems}
IT architecture includes two seemingly incompatible things: on the one hand, it is a large number of peripheral devices with low computing power, low power consumption, high speed of reaction to events, with delay in signal transmission, and on the other hand, cloud servers with high computing power for processing large amounts of data, storing and classifying them, often with elements of machine intelligence and analytics. These two worlds use completely different principles of construction and internal architecture \cite{Nimratz2018}.
\paragraph{Problem identification}
There are several levels of integration at which the Internet of Things and edge technologies pose challenges. The first level is the integration of IoT and Edge with the underlying IT systems deployed in manufacturing, financial, engineering and other areas. Many of these basic systems are obsolete. If they do not have APIs that enable IoT integration, then batch ETL (extract, transform, load) software may be required to load data into these systems.

The second area of ​​challenge is the IoT itself. Many IoT devices are built by independent vendors. They use their own proprietary OS. This makes it difficult to "mix and match" different IoT devices in a single Edge architecture. 
The area of ​​IoT security and compliance is still evolving. Meanwhile, organizations' IT professionals can now ask potential IoT vendors about what is already available for heterogeneous integration and whether they plan to provide interoperability in future products.

\begin{figure}[th]
	\begin{center}
		\includegraphics[scale=0.55]{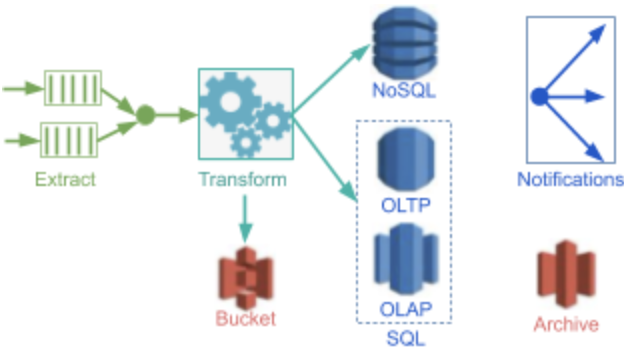}
	\end{center}
	\caption{The typical ETL layer in the IoT sensor network systems}
	\label{fig:example1}
\end{figure}

When it comes to integrating legacy systems, ETL is the one integration method that can help if APIs for the systems are not available. Another alternative is to write an API, but it takes a long time. The good news is that most legacy vendors are aware of the upcoming IoT wave and are already developing their own APIs if they haven't already. IT departments should check with major system vendors to find out what their plans for IoT integration are.
Once the data volume grows significantly, it turns out that building the spacial index itself can be problematic. Either the algorithm goes beyond the definition, or the build time becomes unacceptably long, or the search is ineffective ... As soon as the spacial data begins to change intensively, the spacial index can begin to degrade. Rebuilding the index helps, but not for long.
Etl layer - the level of collection, processing and storage of data.
The back-end ETL (extract, transform, and load) is the third ETL operation. The first was in the peripheral, the second was in the gateway. The back-end ETL collects data from all peripherals and gateways and is responsible for the following operations:
\begin{itemize}
	\item Collection of information,
	\item Bringing information to standard form,
	\item Saving information for future use,
	\item Information lifecycle management including archiving and destruction,
	\item Notifying other services when new data arrives.

\end{itemize}
The Figure~\ref{fig:example1} represents The typical ETL layer in the IoT sensor network systems.

To organize spacial search in multidimensional indices, this work proposes to find a logarithmic way of organizing a computational model. In connection with the problems that arose with R trees \cite{10.1145/971697.602266}, quadrant trees \cite{Muratshin2019}, this work settled on the development of algorithms based on self-similar functions \cite{Muratshin2017}, isomorphic to binary trees\cite{Muratshin2017a}. Self-similar functions \cite{Moon1996} are used to organize multidimensional indexing and searching.

\paragraph{Basic mathematical model} 
Basic mathematical model represents the architechture model for the sensor network system based on publisher/subscriber architecture with the Voronoi diagrams data structure \cite{10.1145/282918.282923} data circulation.
Storage operations (load) are intended for storing, sorting and subsequent retrieval of information. Different tools are used depending on the type of information and its use cases. If the data does not have a strict schema (table columns), then it is stored in NoSQL databases. However, if the data can be systematized with a fixed schema, then SQL database types are used. The latter, in turn, have 2 types - OLTP (Online Transactional Processing) and OLAP (Online Analytic Processing). As the name suggests, the first type is more suitable for the ETL process itself - writing new values ​​to the database, while the second is more convenient for searching and analyzing data. Therefore, often after loading the OLTP database, in the background, the data is copied to OLAP. There are situations when it is not convenient or possible to store data in databases, for example, in the form of a record. This data is written to the data bucket, and the metadata of records is stored in databases. To reduce storage costs, obsolete data is archived or deleted. And the last component of this layer is the internal notification of the presence of new stored data for presentation to clients and for analysis services \cite{Nimratz2019}. This work proposes the NoSQL approach for data processing based on Voronoi diagrams representation of the set of the objects in the system within the construction of the spacial indexation in NoSQL databases.
Every sensor has its own unique identifier ID correspondenting to his IP address. The appriximate model of the traffic exchanges from the sensor to the server proposes the following (\ref{eq:ref1}):

\begin{equation}
\gamma = \frac{\rho\bar{t}}{2(1-\rho)}\frac{\sigma_{a}^2 + \sigma_s^2}{\bar{t}^2}\frac{\bar{t}^2 + \sigma_s^2}{\bar{a}^2+\sigma_s^2} + \bar{t}
 \label{eq:ref1}
\end{equation}
where \(\sigma_a^2, \sigma_s^2\) - variance values of time interval between packets and time
service, \(\bar{a}\) - average value of the interval between packets, \(\bar{t}\) - average service time, \( \rho \) - system's loading.

The flow of messages entering the messaging queue from publishers in the system  (\ref{eq:ref2}):
\begin{equation}
 \varrho(x) = \frac{\exp^{-\lambda} \lambda^x}{x!}
 \label{eq:ref2}
\end{equation}
where \(\varrho(x)\) - possibility of obtaining the \(x\) income signals in the time unit, \(x\) - number of requests per unit of time, \(\lambda\) - the average number of applications per unit of time (rate of receipt of applications), \(exp\) = 2.7182 - base of the natural logarithm.

\section{Methodology}

The main specificity of the Lebesgue's curve \cite{Esculier2006} is localized in the splitting of the subquery.
\begin{itemize}
	\item first we find the starting extent, which is the minimum rectangle that includes the search extent and contains one contiguous range of key values,
	\item calculate the key values,
	\item for the lower left and upper right points of the search extent (KMin, KMax),
	\item find a common binary prefix (from high order to low order) for KMin, KMax,
	\item zeroing all the digits behind the prefix, we get SMin, filling them with ones, we get SMax,
	\item do the inverse transformation of keys to coordinates and find the corners of the starting extent. 
\end{itemize}

\begin{figure}[th]
	\begin{center}
		\includegraphics[width=\textwidth]{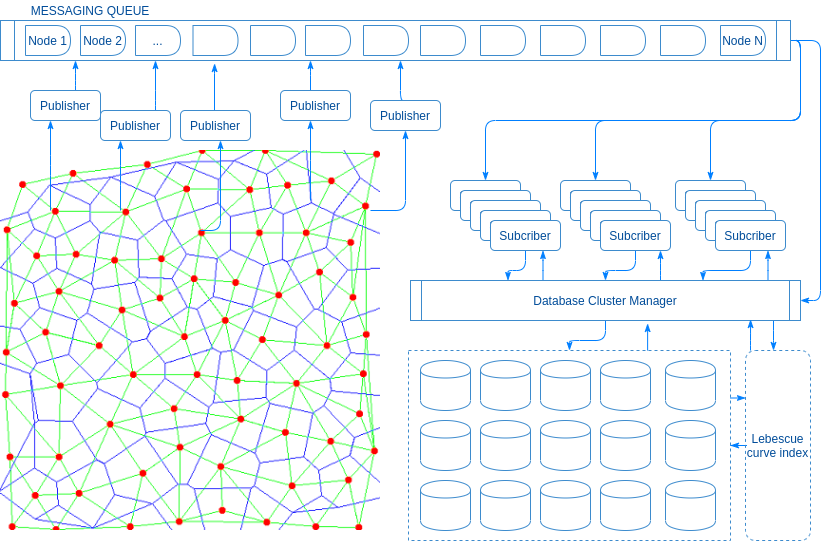}
	\end{center}
	\caption{The proposed transactional model for the ETL level}
	\label{fig:example2}
\end{figure}

\begin{figure}[th]
	\begin{center}
		\includegraphics[width=\textwidth]{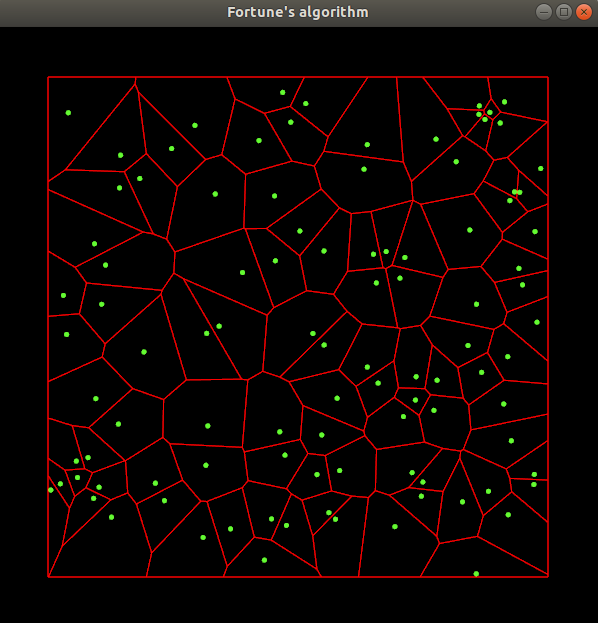}
	\end{center}
	\caption{The static realization of the Voronoi diagrams in 2D}
	\label{fig:example3}
\end{figure}

\begin{figure}[th]
	\begin{center}
		\includegraphics[width=\textwidth]{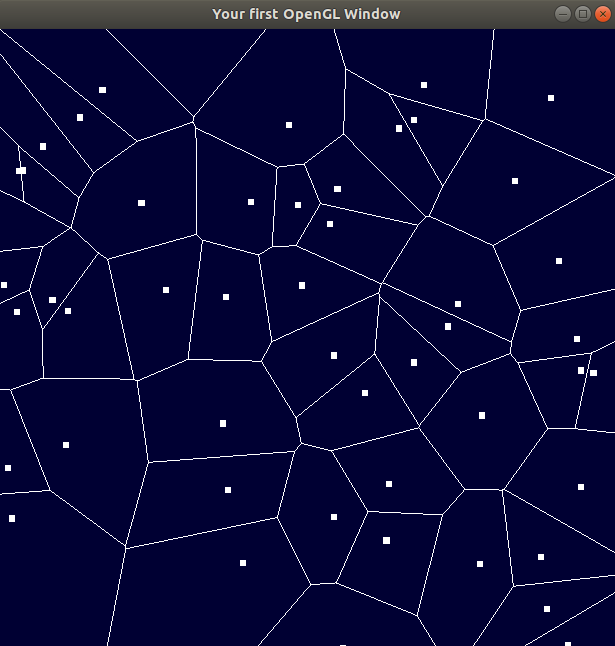}
	\end{center}
	\caption{The dynamic realization of the Voronoi diagrams in 2D}
	\label{fig:example4}
\end{figure}

\begin{figure}[th]
	\begin{center}
		\includegraphics[width=\textwidth]{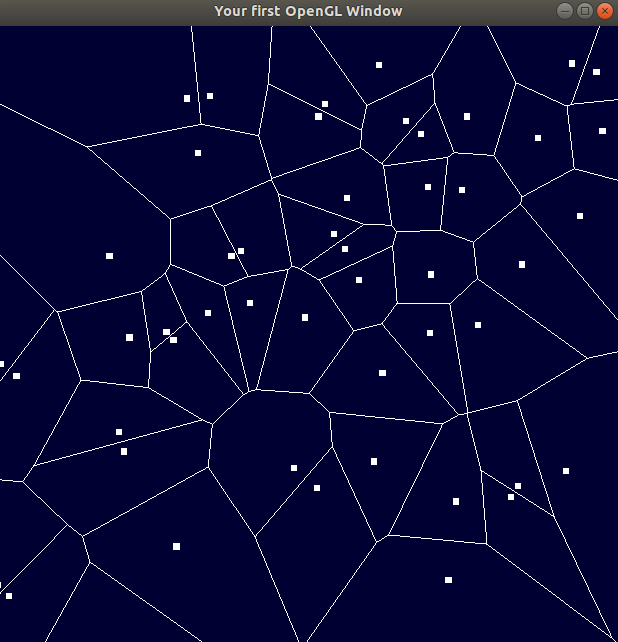}
	\end{center}
	\caption{The static realization of the Voronoi diagrams in 2D}
	\label{fig:example5}
\end{figure}

In the case of the Hilbert curve, by the way, the lower left corner of the starting extent does not necessarily come from SMin, one needs to choose the minimum value. The same with the upper right corner.
The starting extent can be noticeably larger than the search extent, if one is not unlucky, it will turn out to be the maximum extent of the layer (empty prefix).
For the Z-curve, optimization can be done and the starting extent can be equated to the search extent. This is possible due to the peculiarity of the z-curve - for any rectangle, its lower left corner gives the minimum key value, and the upper right corner gives the maximum (in the rectangle). Moreover, such a starting extent can contain more than one range of values, but this is further removed by filtering.
If one pushes the starting extent onto the subquery stack until the subquery stack is empty one will get the top element, if it does not overlap with the search extent, one should discard it and skip iteration. This is in the case the starting extent is larger than the search extent and one needs to ignore the excess.
Furthermore, if the minimum point of the subquery is within the search extent, one queries the index by the value of this point. As a result, one gets two values  - "the first key" higher or equals the desired one and the "last key" on the physical (leaf) page, where the "first key" lies:
\begin{itemize}
	\item if “first key” lower than the maximum value of the current subquery, one should ignore the current subquery, then skip iteration,
	\item if the entire subquery is inside the search extent, one subtracts it by the traverse of the index from the minimum to the maximum value, then end of iteration,
	\item if the “last key” higher than the maximum value of the current subquery, then all the data of the current subquery is on this page, one needs to read the page to the end and filter out the unnecessary, then end of iteration,
	\item the separate case "last key" equials to the maximum value of the current subquery, then one processes separately and traverse forward
	splitting the current subquery,
	\item the one adds 0 and 1 to its prefix - to get two new prefixes,
	\item then one fills in the remainder of the key 0 or 1 - to get the minimum and maximum values ​​of the new subqueries,
	\item one pushes them onto the stack, first the one that added 1, then 0 (this is for unidirectional reading of the index).
\end{itemize}
The Figure~\ref{fig:example2} represents the proposed transactional model for the ETL level.
\paragraph{Results} 
The results represents the first step realizations of the developed transactional ETL model based on Voronoi Diagrams and Lebesgue curves and consists of the realization of the Voronoi diagrams using the C++, Ubuntu OS and OpenGL in 2D dimension in two performances: static version and dynamically recalculating version in time.
The Figure~\ref{fig:example3} represents the static realization of the Voronoi diagrams.
The Figures~\ref{fig:example4} and ~\ref{fig:example5} represents the dynamic realization of the Voronoi diagrams.

Moreover, the results represents the transactional model based on the publisher/subscriber architecture. The Figure~\ref{fig:example2} represents the proposed transactional model for the ETL level.
The static Voronoi Diagrams situates in this repository \hyperref{https://github.com/almakonde/StaticVoronoiDiagrams.git}{algorithm}{the static Voronoi Diagrams}{"the static Voronoi Diagrams realization"}.
The dynamic Voronoi Diagrams situates in this repository \hyperref{https://github.com/almakonde/DynamicVoronoiDiagrams.git}{algorithm}{dynamic Voronoi Diagrams}{"the dynamic Voronoi Diagrams realization"}.
\section{Research}
The research potential of the work lies in the possibility of using algorithms for building maps with the need to plot the density of parameters. The potential lies in the displaying the density maps for the necessary parameters (for example, the dynamics of animals migration, the assessment of the geological resources in different regions, the dynamics of the extinction of wild animal species, the dynamics of the disappearance of forests, the traffic congestion, the consumption of housing and communal resources, the sensitivity of crystals on the surface, the chemical properties of materials, the physical properties of materials, etc...) in loci / zones of the same density.

The research potential of the work lies in the possibility of applying algorithms for constructing multidimensional spacial indexes to optimize search queries in a growing volume of data.
\paragraph{Applications} 
The applications of the proposed model can be realized in multivarious applications in the sensor network systems:
\begin{itemize}
	\item for search operations in the database optimization of the multidimentional spacial geo systems,
	\item for search operations in the distibuted transactions such as blockchain,
	\item for identification of the closest neighbor objects to the object.

\end{itemize}
\paragraph{Implication examples}
The class of the Lebesgue curves (Morton code)\cite{Morton1966}, \cite{Wikipedia2022} can be found in the following database management systems:
\begin{itemize}
	\item AWS database Amazon Aurora \cite{Chandrasekaran2018},
	\item  Amazon RDS for MySQL \cite{Chandrasekaran2018}.

\end{itemize}
\section{Discussion}
Due to the recursive structure of the Voronoi diagrams and Lebesgue curves there exists a possibility to use the recursive defragmentation of the iterations in the both algorithms.
Due to the spacial adaptiveness of the Voronoi diagrams this work proposes to explore the properties of the adaptiveness and time costs of these algorithms in real-time systems. 
The estimated time for Voronoi diagrams calculation is \(N\log{N}\), where \(N\) - is the  number of iterations.
The estimated time of the Lebesque curve indexation also is \(N\log{N}\), where \(N\) - is the  number of iterations.
\section*{References}

\bibliography{mybibfile}

\end{document}